\documentclass[12pt, landscape, twocolumn]{article}

\usepackage{amsmath, amssymb, amsthm}
\usepackage{graphicx}
\usepackage{hyperref}
\usepackage{geometry}
\usepackage{bbm}
\usepackage{cancel}
\usepackage{lineno}
\usepackage{xurl}

\usepackage{tikz}

\usepackage{hyperref}

\usepackage{fancyhdr}
\newtheorem{theorem}{Theorem}[section]

\newcommand{\bR}{\mathbb{R}}
\newcommand{\bZ}{\mathbb{Z}}

\newcommand{\bE}{\mathbb{E}}

\newcommand{\cL}{\mathcal{L}}

\newcommand{\cN}{\mathcal{N}}

\newcommand{\cD}{\mathcal{D}}
\newcommand{\teq}[1]{\stackrel{\text{#1}}{=}}
\newcommand{\din}{d}
\newcommand{\bk}[1]{\langle #1 \rangle}
\newcommand{\bbk}[1]{\big\langle #1 \big\rangle}

\newcommand{\maps}{\text{Maps}}

\newcommand{\Q}[1]{
    \begin{center}
        \vspace{.25cm}
    \begin{minipage}{.45 \textwidth}
        \textbf{Question:} #1
        \vspace{.25cm}
    \end{minipage}
    \end{center}
    }

\newcommand{\C}[1]{
    \begin{center}
        \vspace{.25cm}
    \begin{minipage}{.45 \textwidth}
        #1
    \end{minipage}
    \vspace{.25cm}
    \end{center}
    }

\newcommand{\CT}[2]{
    \begin{center}
        \vspace{.25cm}
    \begin{minipage}{.45 \textwidth}
        \textbf{#1:} #2
        \vspace{.25cm}
    \end{minipage}
    \end{center}
    }

\renewcommand{\O}[1]{
    \begin{center}
        \vspace{.25cm}
    \begin{minipage}{.45 \textwidth}
        \textbf{Observation:} #1
    \end{minipage}
    \vspace{.25cm}
    \end{center}
    }

\title{\vspace{2cm} \bf \Huge TASI Lectures on Physics for Machine Learning}
\author{\LARGE Jim Halverson\vspace{.75cm}\\ \emph{Department of Physics, Northeastern University, Boston, MA 02115, USA}\vspace{.3cm}\\ \emph{The NSF Institute for Artificial Intelligence} \\ \emph{and Fundamental Interactions}\vspace{.3cm} \\ \texttt{jhh@neu.edu}}
\date{}

\begin{document}

\onecolumn

\maketitle

\begin{center}
        \begin{minipage}{0.75\textwidth}
        \begin{abstract}
            These notes are based on lectures I gave at TASI 2024 on Physics for Machine Learning. The focus is on neural network theory, organized according to network expressivity, statistics, and dynamics. I present classic results such as the universal approximation theorem and neural network / Gaussian process correspondence, and also more recent results such as the neural tangent kernel, feature learning with the maximal update parameterization, and Kolmogorov-Arnold networks. The exposition on neural network theory emphasizes a field theoretic perspective familiar to theoretical physicists. I elaborate on connections between the two, including a neural network approach to field theory.
        \end{abstract}
        \end{minipage}

\end{center}

\clearpage
\tableofcontents
\clearpage
\twocolumn

\section{Introduction}

\label{sec:lecture_intro}

Computer science (CS) is still in its infancy.  ``What!?'' you incredulously exclaim, thinking back to the earliest computer scientists you can think of, perhaps Lovelace or Babbage or Turing. Since its early days, CS has steadily progressed, from cracking the Enigma code in the 40's, to personal computers in the 70's and 80's, the internet in the 90's, powerful computers-in-your-pocket in the 2000's, and finally to some version of artificial intelligence in the last decade; see \cite{isaacson2014innovators} for a thorough account. It shows no signs of stopping, and one might wager that future historians will not speak of decades of progress in CS, but centuries. Computer science is still in its infancy.

It is 2024, what have we seen recently? A plethora of ``experimental" breakthroughs in machine learning (ML) that constitute some of the most exciting developments in CS to date. Reinforcement learning algorithms have become world-class at Go \cite{AlphaZero} and Chess \cite{silver2017mastering, lemoine2023alphazero}, merely by playing against themselves. Diffusion models \cite{sohl2015deep,song2020score,ananthaswamy2023physics} can generate images of people that don't exist, and yet you might never know. Large language models (e.g. GPT-3 \cite{GPT3} literature, or popular accounts \cite{roose2022chatgpt,bhatia2023shakespeare}) can code better than you --- or me, at least ---  and they can even write compelling bedtime stories for your children. These forms of artificial intelligence have already started a trillion dollar industry, and some have speculated that we've already seen sparks \cite{bubeck2023sparks}  of AGI, or human-level intelligence.

We've also started to see significant advances in the sciences. For instance, AlphaFold \cite{Jumper2021} has achieved state-of-the-art performance for predicting the structure of folded proteins, and neural networks (NNs) give state-of-the-art models for complex objects, from quantum many-body ground states \cite{carleo2017solving, carrasquilla2017machine} to metrics \cite{Anderson:2020hux,Douglas:2020hpv,Jejjala:2020wcc} on the Calabi-Yau manifolds of string theory. Lack-of-rigor is a natural concern, but in certain cases ML techniques can be made rigorous or interpretable at a level that would satisfy a formal theoretical physicist or pure mathematician; see, e.g., \cite{gukov2024rigor} and references therein for examples in string theory, algebraic geometry, differential geometry, and low-dimensional topology, including uses of ML theory for science.
Of course, this is just scratching the surface, see, e.g., \cite{IAIFI2023,MLPS2023,Carleo:2019ptp} for a smattering of results across physics and the physical sciences.

With this in mind, we have some explaining to do. There are a large number of impressive ML results in both pop culture and the sciences, but experiments have significantly outpaced theory over the last decade. An analogy from physics might be useful, given the original audience. The current situation in ML might be reminiscent of the 1960's in particle physics: some essentials of the theory were known, e.g. quantum electrodynamics (QED), but a deluge of newly discovered hadrons left theorists scratching their heads and demanded an organizing principle to explain the new phenomena. This was one of the motivating factors for the development of string theory, but the right theory in the end is the quark model, as described by quantum chromodynamics (QCD). It's tempting to speculate that recent results in ML are waiting for someone to discover a foundational theory, akin to QCD for the strong interactions. As theoretical physicists, we hope for this sort of mechanistic understanding. However, it might be that ML and intelligence are so complex that sociology is a better analogy, and it would be a fool's errand to expect such a precise theory. My personal guess is that detailed theory will pay off in certain areas of ML --- perhaps related to sparsity in NNs, architecture design, or the topics of these lectures --- but not in all. Nevertheless, given the experimental progress, one must try.

\bigskip
These lectures were given at TASI 2024 \cite{TASI2024} to Ph.D. students in theoretical high energy physics. I was given the title ``Physics4ML", which means that I'm trying to keep ML-for-Physics to a minimum. I take a physics perspective, but acknowledge that this is an enormous field to which many communities have contributed.

The central idea of the lectures is that understanding NNs is essential to understanding ML, a story I built on three pillars: expressivity, statistics, and dynamics of NNs, one for each of the lectures I gave. The content builds on many ideas explained to me over the last five years by friends, acknowledged below, who pioneered many of the developments, as well as a few of my own results. Throughout, I take a decidedly field-theoretic lens to suit the audience, and also because I think it's a useful way to understand NNs.

\clearpage
\subsection*{The Setup.}
Understanding ML at the very least means understanding neural networks. A neural network is a function
\begin{equation}
\phi_\theta:\bR^d \to \bR
\end{equation}
with parameters $\theta$. We've chosen outputs in $\bR$ because, channeling Coleman, scalars already exhibit the essentials. We'll use the lingo
\begin{align}
    \text{Input:} & \quad x \in \bR^d \\
    \text{Output:} & \quad \phi_\theta(x) \in \bR \\
    \text{Network:} & \quad \phi_\theta \in \maps(\bR^d, \bR) \\
    \text{Data:} & \quad \cD,
\end{align}
where the data $\cD$ depends on the problem, but involves at least a subset of $\bR^d$, potentially paired 
with labels $y\in \bR$.

With this minimal background, let's ask our central question:
\Q{What does a NN predict?}
For any fixed value of $\theta$, the answer is clear: $\phi_\theta(x)$. However, the answer is complicated by issues of both dynamics and statistics. 

First, dynamics. In ML, parameters are updated to solve problems and we really have {\bf trajectories} in 
\begin{align}
    \text{Parameter Space:} & \quad \theta(t) \in \bR^{| \theta |} \\
    \text{Output Space:} & \quad \phi_{\theta(t)}(x) \in \bR \\ 
    \text{Function Space:} & \quad \phi_{\theta(t)}\in \maps(\bR^d, \bR).
\end{align}
governed by some learning dynamics determined by the optimization algorithm and the nature of the learning problem. For instance, in supervised learning we have data
\begin{equation}
\cD = \{(x_\alpha, y_\alpha) \in \bR^d \times \bR\}_{\alpha=1}^{|\cD|},
\end{equation}
and a loss function 
\begin{equation}
\cL[\phi_\theta] = \sum_{\alpha=1}^{|\cD|} \ell(\phi_\theta(x_\alpha), y_\alpha),
\end{equation}
where $\ell$ is a loss function such as $\ell_{\text{MSE}} = (\phi_\theta(x_\alpha)-y_\alpha)^2$. One may optimize $\theta$ by gradient descent 
\begin{equation}
    \frac{d\theta_i}{dt} = -\nabla_{\theta_i} \cL[\phi_\theta],
\end{equation}
or other algorithms, e.g., classics like stochastic gradient descent (SGD) \cite{sgd,perceptron} or Adam \cite{kingma2017adammethodstochasticoptimization}, or a more recent technique such as Energy Conserving Descent \cite{de2022born,de2023improving}. Throughout, $t$ is training time of the learning algorithm unless otherwise noted.

Second, statistics. When a NN is initialized on your computer, the parameters $\theta$ are initialized as draws
\begin{equation}
    \theta \sim P(\theta)
\end{equation}
from a distribution $P(\theta)$, where $~$ means ``drawn from" in this context.
Different draws of $\theta$ will give different functions $\phi_\theta$, and a priori we have no reason to prefer one over another. The prediction $\phi_\theta(x)$ therefore can't be fundamental! Instead, what is fundamental is the average prediction and second moment or variance:
\begin{align}
    \bE[\phi_\theta(x)] & = \int d\theta P(\theta) \, \phi_\theta(x) \\
    \bE[\phi_\theta(x) \phi_{\theta}(y)] & = \int d\theta  P(\theta)  \, \phi_\theta(x) \phi_{\theta}(y),
\end{align}
as well as the higher moments. Expectations are across different initializations. Since we're physicists, we henceforth replace $\bE[\cdot]=\bk{\cdot}$ and we remember this is a statistical expectation value. It's useful to put this in our language:
\begin{align}
    G^{(1)}(x) & = \bk{\phi_\theta(x)} \\
    G^{(2)}(x, y) & = \bk{\phi_\theta(x) \phi_\theta(y)},
\end{align}
the mean prediction and second moment are just the one-point and two-point correlation functions of the statistical ensemble of neural networks. Apparently ML has something to do with field theory. 

Putting the dynamics and statistics together, we have an ensemble of initial $\theta$-values, each of which is the starting point of a trajectory $\theta(t)$, and therefore we have an ensemble of trajectories. We choose to think of $\theta(t)$ drawn as 
\begin{equation}
    \theta(t) \sim P(\theta(t)),
\end{equation}
a density on parameters that depends on the training time and yields time-dependent correlators 
\begin{align}
    G_t^{(1)}(x) & = \bk{\phi_\theta(x)}_t \\
    G_t^{(2)}(x, y) & = \bk{\phi_\theta(x) \phi_\theta(y)}_t,
\end{align}
where the subscript $t$ indicates time-dependence and the expectation is with respect to $P(\theta(t))$. Of course, assuming that learning is helping, we wish to take $t\to\infty$ and are interested in
\C{$G_\infty^{(1)}(x)$ = mean prediction of $\infty$-number of NNs as $t\to\infty$.}
Remarkably, we will see that in a certain supervised setting there is an exact analytic solution for this quantity.

There is one more pillar beyond dynamics and statistics that I need to introduce: expressivity. Let's make the point by seeing a failure mode. Consider neural networks of the following functional form, or \emph{architecture}:
\begin{equation}
\phi_\theta(x) = \theta \cdot x.
\end{equation}
The one-point and two-point functions of these are analytically solvable, but going that far defeats the purpose: this is a linear model, and learning schemes involving it are linear regression, which will fail on general problems. We say that model is not \emph{expressive} enough to account for the data. Conversely, we wish instead to choose \emph{expressive} architectures that can model the data, which essentially requires that the architecture is complex enough to approximate anything. Of course, such architectures must be non-linear.

We now have the three {\bf pillars} on which we'll build our understanding of neural networks, and associated questions:
\begin{itemize}
    \item {\bf Expressivity.} How powerful is the NN?
    \item {\bf Statistics.} What is the NN ensemble?
    \item {\bf Dynamics.} How does it evolve?
\end{itemize}
We will approach the topics in this order, building up theory that takes us through the last few years through a physicist's lens. 

A physicist's lens means a few things. First, it means a physicist's tools, including:
\begin{itemize}
\item {\bf Field Theory} as alluded to above.
\item {\bf Landscape Dynamics} from loss functions on $\theta$-space.
\item {\bf Symmetries} of individual NNs and their ensembles.
\end{itemize} 
A physicist's lens also means we will try to find the appropriate balance of rigor and intuition. We'd like mechanisms and toy models, but we won't try to prove everything at a level that would satisfy a mathematician. This befits the current situation in ML, where there are an enormous number of empirical results that need an $O(1)$ theoretical understanding, not rigorous proof.

Henceforth, we drop the subscript $\theta$ in $\phi_\theta(x)$, and the reader should recall that the network depends on parameters.

\section{Expressivity of Neural Networks}
\label{sec:expressivity}

Neural networks are big functions composed out of many simpler functions according to the choice of architecture. The compositions give more flexibility, prompting
\Q{How powerful is a NN?}
Ultimately this is a question for mathematics, as it's a question about functions.  Less colloquially, what we mean by the power of a NN is its ability to approximate any function.

\subsection{Universal Approximation Theorem}

The Universal Approximation Theorem (UAT) is the first result in this direction. It states that a neural network with a single hidden layer can approximate any continuous function on a compact domain to arbitrary accuracy. More precisely, the origin version of the theorem due to Cybenko is
\begin{theorem}[Cybenko]
Let $f: \bR^d \to \bR$ be a continuous function on a compact set $K\subset \bR^d$. Then for any $\epsilon>0$ there exists a neural network with a single hidden layer of the form
\begin{equation}
\phi(x) = \sum_{i=1}^N \sum_{j=1}^d w_i^{(1)} \sigma(w_{ij}^{(0)}  x_j + b^{(0)}_i)+b^{(1)}, \label{eqn:perceptron}
\end{equation}
$\theta = \{w_{ij}^{(0)}, w_i^{(1)}, b^{(0)}_i, b^{(1)}\}$,
where $\sigma:\bR \to \bR$ is a non-polynomial non-linear activation function, such that
\begin{equation}
\sup_{x\in K} |f(x) - \phi(x)| < \epsilon.
\end{equation}
\end{theorem}
\noindent The architecture in Eq.~\eqref{eqn:perceptron} is known by many names, including the perceptron, a fully-connected network, or a {\bf feedforward network}. The parameter $N$ is known as the {\bf width}. It may be generalized to include a {\bf depth} dimension $L$ encoding the number of compositions of affine transformations and non-linearities. Such an architecture is known as a \emph{multi-layer} perceptron (MLP) or a deep feedforward network.

The UAT is a powerful result, but it has some limitations. First, though the error $\epsilon$ does get better with $N$, it doesn't say how many neurons are needed. Second, it doesn't say how to train the network: though there is a point $\theta^*$ in parameter space that is a good approximation to any $f$, existence doesn't imply that we can find it, which is a question of learning dynamics.

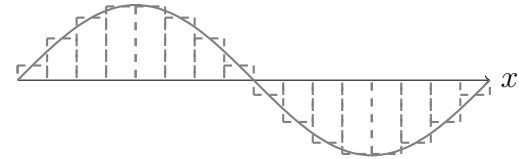
\begin{figure}
    \centering
\begin{tikzpicture}

    \draw[->] (0,0) -- (2*pi,0) node[right] {$x$};

    \draw[domain=0:2*pi, smooth, variable=\x, gray, thick] 
        plot ({\x}, {sin(\x r)});
    
    \foreach \i in {0,...,15}
    {
        \pgfmathsetmacro{\xstart}{\i * pi / 8}
        \pgfmathsetmacro{\xend}{(\i + 1) * pi / 8}
        \pgfmathsetmacro{\xmid}{(\xstart + \xend) / 2}
        \pgfmathsetmacro{\ymid}{sin(\xmid r)}
        \draw[gray, dashed, thick] (\xstart,\ymid) -- (\xend,\ymid);
    }

    \foreach \i in {1,...,15}
    {
        \pgfmathsetmacro{\xpos}{\i * pi / 8}
        \pgfmathsetmacro{\xprev}{(\i - 1) * pi / 8}
        \pgfmathsetmacro{\xnext}{\i * pi / 8}
        \pgfmathsetmacro{\ymid}{sin((\xpos - pi / 16) r)}
        \pgfmathsetmacro{\yprev}{sin((\xprev + \xnext) / 2 r)}
        \draw[gray, dashed, thick] (\xpos, \yprev) -- (\xpos,0);
        \draw[gray, dashed, thick] (\xpos - pi/8, \yprev) -- (\xpos - pi/8,0);
    }

    \pgfmathsetmacro{\yinit}{sin((0 + pi/8) / 2 r)}
    \pgfmathsetmacro{\yfinal}{sin((15*pi/8 + 2*pi) / 2 r)}
    \draw[gray, dashed, thick] (0, \yinit) -- (0, 0);
    \draw[gray, dashed, thick] (2*pi, \yfinal) -- (2*pi, 0);


\end{tikzpicture}
\label{fig:UAT_sin}
\caption{The Universal Approximation Theorem can be understood by approximating a function like $\sin(x)$ with a series of bumps.}
\end{figure}

To ask the obvious,
\Q{Why does this UAT work?} 
In Cybenko's original work, he focused on the case that $\sigma$ was the sigmoid function 
\begin{equation}
\sigma(x) = \frac{1}{1+e^{-x}},
\end{equation}
which allows us to get a picture of what's happening. The sigmoids appear in the network function of the form 
\begin{equation}
\sigma\left(w_i^{(0)}\cdot x + b_i^{(0)}\right)
\end{equation}
which approximates a shifted step function as $w^{(0)}\to\infty$. A linear combination can turn it into a bump with approximately compact support that gets scaled by $w^{(1)}$. These bumps can then be put together to approximate any function; see, e.g., Fig.~\ref{fig:UAT_sin}.

Cybenko's version of the UAT was just the first, and there are many generalizations, including to deeper networks, to other activation functions, and to other domains.  See \cite{cybenko1989approximation} for the original paper and \cite{hornik1991approximation} for a generalization to the case of multiple hidden layers.

\subsection{Kolmogorov-Arnold Theorem}
\begin{figure}[th]
    \begin{center}
    \includegraphics[width=1.05\columnwidth]{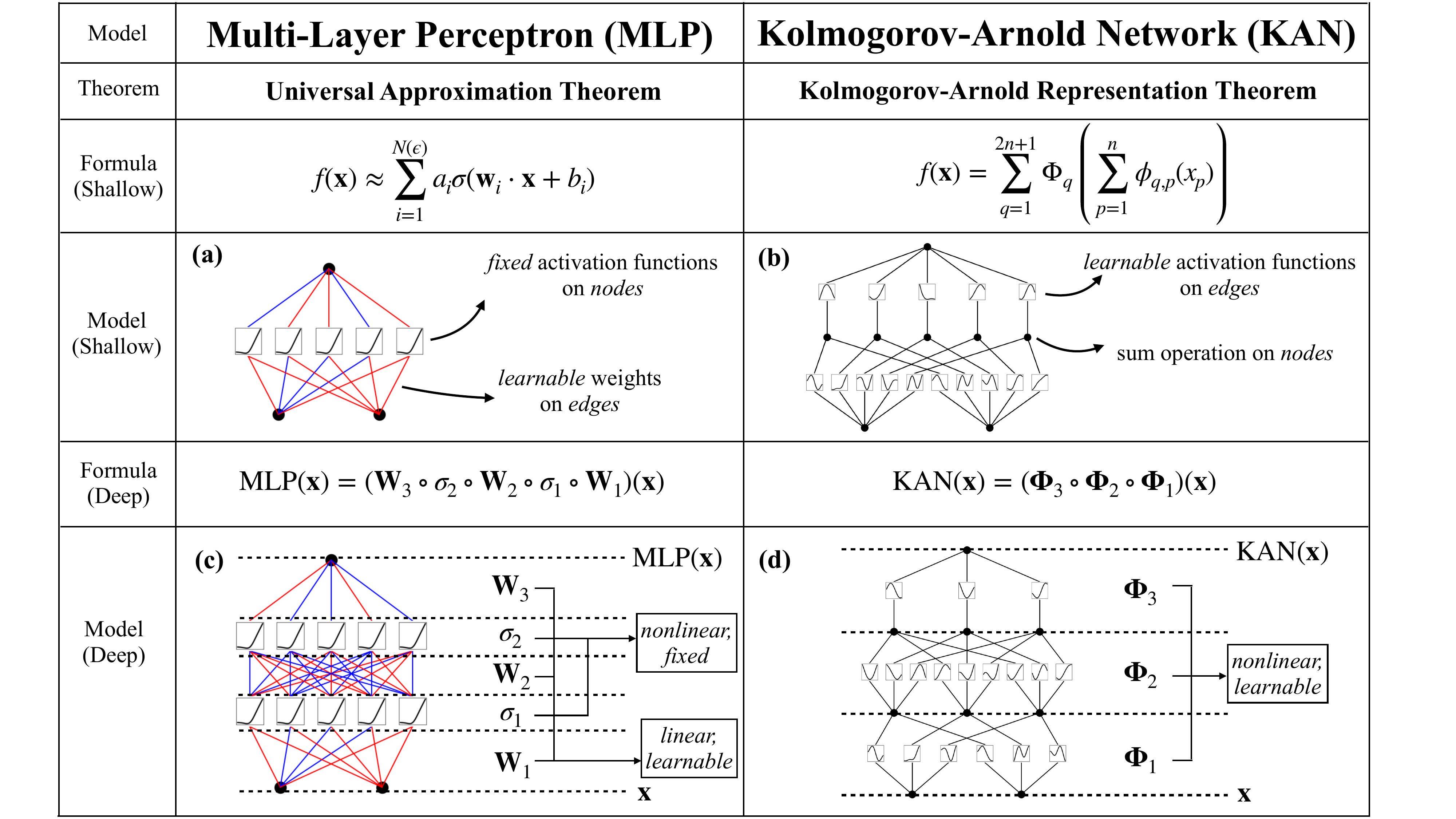}
    \end{center}
    \caption{A comparison of MLP and KAN, including their functional form and the mathematical theorem motivating the architecture.}
    \label{fig:KAN-figure}
    \end{figure}
The UAT essentially says that any function can be approximated by a neural network, i.e. a complicated function comprised out of many building blocks. It is natural to wonder whether there are other theorems to this effect, and whether they inspire other neural network constructions.

One such result is due to Kolmogorov \cite{kolmogorov1957representation} and Arnold \cite{arnold1957representation}, who showed that any multivariate continuous function can be represented (exactly, not approximated) as a sum of continuous one-dimensional functions:

\begin{theorem}[Kolmogorov-Arnold Representation Theorem]
Let $f:[0,1]^n\to \bR$ be an arbitrary multivariate continuous function. Then it has the representation 
\begin{equation}
    f(x_1,\dots,x_n) = \sum_{q=0}^{2n}\Phi_q\left(\sum_{p=1}^n \phi_{q,p}(x_p)\right) \label{eqn:KART}
\end{equation}
with continuous one-dimensional functions $\Phi_q$ and $\phi_{q,p}$.
\end{theorem}

This result is perhaps less intuitive than the UAT. If every multivariate function is to be represented by univariate functions plus addition, what then of $f(x,y)=xy$? This is just a pedagogical example since
\begin{equation}
    f(x,y) = xy = e^{\log x + \log y},
\end{equation}
but of course the general theorem is very non-trivial.

A notable feature of \eqref{eqn:KART} is that it is a sum of functions of functions, which one might take to inspire a neural network definition. However, one aspect that is a little unusual is that $\phi_{q,p}$ depends on both $p$ and $q$, and therefore lives on the connection between $x_p$ and $x^{(1)}_{q}:= \sum_{p=1}^n \phi_{q,p}(x_p)$. If this is a neural network, it's one with activations $\phi_{q,p}$ \emph{on the edges}, rather than on the nodes. Furthermore, for distinct values of $p,q$, the functions $\phi_{q,p}$ are independent in general.

In fact, this idea led to the recent proposal of Kolmogorov-Arnold networks (KAN) \cite{KAN}, a new architecture in which activation functions are on the edges, as motivated by KART, and are learned. Much like how MLPs may be formed out of layers that repeat the structure in the single-layer UAT, KANs may be formed into layers to repeat the structure of KART. See Figure \ref{fig:KAN-figure} for a detailed comparison with MLPs, including a depiction of the architectures and how they are formed out of layers motivated by their respective mathematical theorems. The KAN implementation provided in \cite{KAN} includes visualizations of the learned activation functions, which can lead to interpretable can architectures that can be mapped directly onto symbolic formulae.

\section{Statistics of Neural Networks}
\label{sec:statistics}

Let's try to understand neural networks at initialization. 
 For this, firing up your computer once is not enough, since one initialization give you one draw $\theta\sim P(\theta)$ and therefore one random neural network. Instead, we want to understand the \emph{statistics} of the networks:
 \Q{What characterizes the stats of the NN ensemble?}
 One aspect of this is encoded in the moments, or $n$-pt functions,
\begin{equation}
G^{(n)}(x_1,\dots,x_n) = \bk{\phi(x_1)\dots \phi(x_n)},
\end{equation}
which may be obtained from a partition function as 
\begin{align}
Z[J] &= \bk{e^{\int d^dx J(x) \phi(x)}} \\
G^{(n)}(x_1,\dots,x_n) &= \left(\frac{\delta}{\delta J(x_1)}\dots \frac{\delta}{\delta J(x_n)} Z[J]\right)\bigg|_{J=0},
\end{align}
where $J(x)$ is a source.
This expectation $\bk{\cdot}$ is intentionally not specified here to allow for flexibility. For instance, using the expectation in the introduction we have
\begin{align}
    \label{eqn:Zparam}
    Z[J] & = \int d\theta P(\theta) e^{\int d^dx  J(x) \phi(x)},
\end{align}
reminding the reader that the NN $\phi(x)$ depends on $\theta$. The partition function integrates over the density of network parameters. But as physicists we're much more familiar with function space densities according to 
\begin{align}
    \label{eqn:Zfunc}
Z[J] & = \int \cD\phi \,e^{-S[\phi]} e^{\int J(x) \phi(x)},
\end{align}
the Feynman path integral that determines the correlators from an action $S[\phi]$ that defines a density on functions.

Since starting a neural network requires specifying the data $(\phi,P(\theta))$, the \emph{parameter space partition function} \eqref{eqn:Zparam} and associated parameter space calculation of correlators is always available to us. Given that mathematical data, one might ask
\Q{What is the action $S[\phi]$ associated to $(\phi,P(\theta))$?}
When this question can be answered, it opens a second way of studying or understanding the theory. The parameter-space and function-space descriptions should be thought of as a \emph{duality}.

\subsection{NNGP Correspondence}

Having raised the question of the action $S[\phi]$ associated to the network data $(\phi,P(\theta))$, we can turn to a classic result of Neal \cite{neal}.

For simplicity, we again consider a single-layer fully connected network of width $N$, with the so-called biases turned off for simplicity:
\begin{equation}
    \phi(x) = \frac{1}{\sqrt{N}}\sum_{i=1}^N \sum_{j=1}^d w_i^{(1)} \sigma(w_{ij}^{(0)} x_j), \label{eqn:biasless_perceptron}
    \end{equation}
where the set of network parameters is $\theta = \{w_{ij}^{(0)}, w_i^{(1)}\}$ independently and identically distributed (i.i.d.). 
\begin{equation}
    w_{ij}^{(0)}\sim P(w^{(0)}) \qquad w_i^{(1)} \sim P(w^{(1)}).
\end{equation} 
Under this assumption, we see
\O{The network is a sum of $N$ i.i.d. functions.}
This is a function version of the Central Limit Theorem, generalizing the review in Appendix \ref{app:CLT}, and gives us the Neural Network / Gaussian Process (NNGP) correspondence, 
\CT{NNGP Correspondence}{in the $N \to \infty$ limit, $\phi$ is drawn from a Gaussian Process (GP), 
\begin{equation}
    \lim_{N\to \infty} \,\,\,\phi(x) \sim \cN\left(\mu(x), K(x,y)\right),
\end{equation}
with mean and covariance (or kernel) $\mu(x)$ and $K(x,y)$.}
By the CLT, $\exp(-S[\phi])$ is Gaussian and therefore
$S[\phi]$ is quadratic in networks. Now this really feels like physics, since the infinite neural network is drawn from a Gaussian density on functions, which defines a generalized free field theory.

We will address generality of the NNGP correspondence momentarily, but let's first get a feel for how to do computations.
To facilitate then, we take $P(w^{(1)})$ to have zero mean and finite variance, 
\begin{equation}
    \bk{w^{(1)}}=0 \qquad  \bk{w^{(1)}w^{(1)}} = \mu_2,
\end{equation} which causes the one-point function to vanish
$G^{(1)}(x) = 0.$ Following Williams \cite{williams}, we compute the two-point function in parameter space (with Einstein summation)
\begin{align}
    G^{(2)}(x,y) 
        & = \frac{1}{N}\bk{w_i^{(1)} \sigma(w_{ij}^{(0)} x_j)\,\,w_k^{(1)}  \sigma(w_{kl}^{(0)} y_l)} \\
        & \teq{} \frac{1}{N}\, \bk{w_i^{(1)} w_k^{(1)}} \bk{\sigma(w_{ij}^{(0)} x_j) \sigma(w_{kl}^{(0)} y_l)} \\
        & = \frac{\mu_2}{N} \bk{\sigma(w_{ij}^{(0)} x_j)\,  \sigma(w_{il}^{(0)} y_l)},
\end{align}
where the last equality follows from the ones being i.i.d., $\bk{w_i^{(1)}w_k^{(1)}} = \mu_2 \delta_{ik}$. The sum over $i$ gives us $N$ copies of the same function, leaving us with 
\begin{equation}
    G^{(2)}(x,y) = \mu_2\,\, \bk{\sigma(w_{ij}^{(0)} x_j)\,  \sigma(w_{il}^{(0)} y_l)},
\end{equation}
where we emphasize there is now 
\emph{no summation on $i$}. This is an exact-in-$N$ two-point function that now requires only on the computation of the quantity in bra-kets. One may try to evaluate it exactly by doing the integral over $w^{(0)}.$ If it can't be done, Monte Carlo estimates may be obtained from $M$ samples of $w^{(0)}\sim P(w^{(0)})$ as
\begin{equation}
    G^{(2)}(x,y) \simeq \frac{\mu_2}{M}\sum_{\text{samples}}^M \sigma(w_{ij}^{(0)} x_j)\,  \sigma(w_{il}^{(0)} y_l).
\end{equation}
In typical NN settings, parameter densities are easy to sample for convenience, allowing for easy computation of the estimate. If the density is more complicated, one may always resort to Markov chains, e.g. as in lattice field theory.

With this computation in hand, we have the defining data of this NNGP,
\begin{equation}
    \lim_{N\to \infty} \,\,\,\phi(x) \sim \cN\left(0, G^{(2)}(x,y)\right).
\end{equation}
The associated action is 
\begin{equation}
S[\phi] = \int d^dx d^dy\, \phi(x) \,G^{(2)}(x,y)^{-1}\, \phi(y),
\end{equation}
where 
\begin{equation}
\int d^dy \, G^{(2)}(x,y)^{-1} G^{(2)}(y,z) = \delta^{(d)}(x-z).
\end{equation}
defines the inverse two-point function. In fact, this allows us to determine the action of any NNGP with $\mu(x)=G^{(1)}(x)=0$, by computing the $G^{(2)}$ in parameter space and inverting it.

\vspace{.5cm}
So certain large neural networks are function draws from generalized free field theories. But at this point you might be asking yourself
\Q{How general is the NNGP correspondence?}
Neal's result --- that infinite-width single-layer feedforward NNs are drawn from GP --- stood on its own for many years, perhaps (I am guessing) due to focus on non-NN ML techniques in the 90's and early 2000's during a so-called AI Winter. As NNs succeeded on many tasks in the 2010's after AlexNet \cite{NIPS2012_c399862d}, however, many asked whether architecture $X$ has a hyperparameter $N$ such that the network is drawn from a Gaussian Process as $N\to \infty$. Before listing such $X$'s, let's rhetorically ask
\Q{Didn't Neal's result essentially follow from summing $N$ i.i.d. random functions? Maybe NNs do this all the time?}
In fact, that is the case. Architectures admitting an NNGP limit include
\begin{itemize}
    \item \textbf{Deep Fully Connected Networks,} $N=\text{width}$,
    \item \textbf{Convolutional Neural Networks,} $N=\text{channels}$,
    \item \textbf{Attention Networks,} $N=\text{heads}$,
\end{itemize}
and many more. See, e.g., \cite{yang2019wide} and references therein.

\subsection{Non-Gaussian Processes}

If the GP limit exists due to the CLT, then violating any of the assumptions of the CLT should introduce non-Gaussianities, which are interactions in field theory. From Appendix \ref{app:CLT}, we see that the CLT is violated by finite-$N$ corrections and breaking statistical independence. See \cite{Demirtas:2023fir} for a systematic treatment of independence breaking, and derivation of NN actions from correlators.

We wish to see the $N$-dependence of the connected $4$-pt function. William's technique for computing $G^{(2)}$ extends to any correlator. To avoid a proliferation of indices, we will compute it using the notation
\begin{equation}
\phi(x) = \sum_i w_i \varphi_i(x)
\end{equation}
where $w_i$ is distributed as $w^{(1)}$ was in the single layer case, and $\varphi_i(x)$ are i.i.d. neurons of \emph{any} architecture. The four-point function is 
\begin{align}
G^{(4)} &= \bk{\phi(x)\phi(y)\phi(z)\phi(w)} \\ 
    &= \sum_{i,j,k,l} \bk{w_i w_j w_k w_l} \bk{\varphi_i(x)\varphi_j(y)\varphi_k(z)\varphi_l(w)} \\ 
    &= \sum_i \bk{w_i^4} \bk{\varphi_i(x)\varphi_i(y)\varphi_i(z)\varphi_i(w)} \\ &\,\,\,\,\,\,\,\,+ \sum_{i\neq j} \bk{w_i^2} \bk{w_j^2}\bk{\varphi_i(x)\varphi_i(y)\varphi_j(z)\varphi_j(w) + \text{perms}}.
\end{align}
One can see that you have to be careful with indices. The connected $4$-pt function is \cite{yaida2020non}
\begin{equation}
G^{(4)}_c(x_,y,z,w) = G^{(4)}(x,y,z,w) - \left(G^{(2)}(x,y)G^{(2)}(z,w) + \text{perms}\right),
\end{equation}
and watching indices carefully we obtain 
\begin{align}
G^{(4)}_c(x_,y,z,w) = \frac{1}{N} \bigg(\mu_4 \, \bbk{\varphi_i(x)\varphi_i(y)\varphi_i(z)\varphi_i(w)} \\ - \mu_2^2 \,\left(\bbk{\varphi_i(x)\varphi_i(y)}\bk{\varphi_i(z)\varphi_i(w)} + \text{perms}\right)\bigg),
\end{align}
with no Einstein summation on $i$.
We see that the connected $4$-pt function is non-zero at finite-$N$, signalling interactions. We will see that in some examples $G^{(4)}_c$ can be computed exactly.

\subsection{Symmetries}

We have gotten some control over the statistics of the ensemble of neural networks. It natural 
at this point to ask 
\Q{Is there structure in the ensemble?}
By this I mean properties that the ensemble realizes that an individual network might not see. We will return to this question broadly in Section \ref{sec:NNFT}, but for now we will focus on one type of structure: symmetries.

To allow for symmetries at both input and output, in this section we consider networks
\begin{equation}
\phi: \bR^d \to \bR^D.
\end{equation} 
with $D$-dimensional output. Sometimes the indices will be implicit.

A classic example is equivariant neural networks \cite{pmlr-v48-cohenc16}.
We say that $\phi$ is \textbf{$G$-equivariant} with respect to a group $G$ if 
\begin{equation}
\rho_{D}(g) \phi(x) = \phi(\rho_d(g) x) \qquad \forall g \in G,
\end{equation}
where 
\begin{equation}
    \rho_d \in \text{Mat}(\bR^d)\qquad \rho_D \in \text{Mat}(\bR^D)
\end{equation}
are matrix representations of $G$ on $\bR^D$ and $\bR^d$, respectively. The network is \textbf{invariant} if $\rho_D = \mathbbm{1}$, the trivial representation.
Equivariance is a powerful constraint on the network that may be implemented in a variety of ways. For problems that have appropriate symmetries, building them into the network can improve the speed and performance of learning \cite{winkels20183d}, e.g., at the level of scaling laws \cite{kaplan2020scaling, Batzner2022,frey2022neural}. For instance, in Phiala Shanahan's lectures you'll learn about $SU(N)$-equivariant neural networks from her work \cite{boyda2021sampling}, which are natural in lattice field theory due to invariance of the action under gauge transformations.

But this is the statistics section, so 
we're interested in symmetries that arise in \emph{ensembles} of neural networks, which leave the statistical ensemble invariant. In field theory, we call them \textbf{global symmetries}. Let the network transform under a group action as 
\begin{equation}
\phi \mapsto \phi_g, \qquad g \in G.
\end{equation}
We say that ensemble of networks has a global symmetry group $G$ if the partition function is invariant,
\begin{equation}
    Z_g[J] = Z[J], \qquad \forall g \in G.
\end{equation}
At the level of expectations, this is 
\begin{equation}
 \bk{e^{\int d^d x J(x) \phi_g(x)}}  =\bk{e^{\int d^d x J(x) \phi(x)}}  \qquad \forall g \in G,
\end{equation}
where one can put indices on $\phi$ and $J$ as required by $D$. By a network redefinition on the LHS, this may be cast as having a symmetry if $\bk{\cdot}$ is invariant. 
In the usual path integral this is the statement of invariant action $S[\phi]$ and measure $\cD\phi$. In parameter space, the redefinition may be instituted \cite{Maiti:2021fpy} by absorbing $g$ into a redefinition of parameters as $\theta \mapsto \theta_g$, with symmetry arising when 
\begin{equation}
\int d\theta_g P(\theta_g) e^{\int d^d x J(x)\phi_\theta(x)} = \int d\theta P(\theta) e^{\int d^d x J(x)\phi_\theta(x)},
\end{equation}
i.e. the parameter density and measure must be invariant. We will give a simple example realizing this mechanism in a moment.

It is most natural to transform the input or output of the network. Our mechanism allows for symmetries of both types, which are analogs of spacetime and internal symmetries, respectively. It may also be interesting to study symmetries of intermediate layers, if one wishes to impose symmetries on learned representations. Equivariance fits into this picture because it turns a transformation at input into a transformation at output. The ensemble of equivariant NNs is invariant under $\rho_d$ action on the input if the partition function is invariant under the induced $\rho_D$ action on the output.

\medskip
\noindent \textbf{Example.} \label{sec:stats_examples} Consider any architecture of the form 
\begin{equation}
    \phi(x) = \sum_i w_i \varphi_i(x), \qquad w_i \sim P(w) \, \text{even},
\end{equation}
for \emph{any} neuron $\varphi$, which could itself be a deep network. The $\bZ_2$ action 
$\phi \mapsto -\phi$ may be absorbed into parameters $w_g = -w_i$ with $dw P(w)$ invariant by evenness, which is a global symmetry provided that the domain is invariant. This theory has only even correlators. 

Less trivial examples will be presented below, when we present specific networks and compute correlators.

\subsection{Examples}

In the statistics of neural networks, we have covered three topics: Gaussian limits, non-Gaussian corrections from violating CLT assumptions, and symmetries. We will present this essential data in some examples and then discuss similarities and differences. For canonical cases in ML, like networks with ReLU activation or similar, see \cite{Halverson:2020trp} and references therein.

\medskip
\noindent \emph{Gauss-net.}
The architecture and parameter densities are 
\begin{align}
\phi(x) = \frac{w^{(1)}_{i} \exp(w^{(0)}_{ij} x_{j} + b^{(0)}_{i})}{\sqrt{\exp[2(\sigma_{b_0}^2 + \frac{\sigma^2_{w_0}}{\din}x^2 )]}},
\end{align}
for parameters drawn i.i.d. from 
\begin{equation}
    w^{(0)} \sim \mathcal{N}(0,\frac{ \sigma^2_{W_0}}{\din}),\qquad w^{(1)} \sim \mathcal{N}(0, \frac{\sigma^2_{w_1}}{2N}),\qquad b^{(0)} \sim \mathcal{N}(0, \sigma^2_{b_0}).
\end{equation}
The two-point function is
\begin{equation}
    G^{(2)}(x_1, x_2) = \frac{\sigma_{w_1}^2}{2} e^{-\frac{1}{2\din}\sigma_{w_0}^{2} (x_1-x_2)^2}
\end{equation}
and we see that the theory has a correlation length 
\begin{equation}
\xi = \sqrt{\frac{\din}{\sigma_{w_0}^2}}.
\end{equation}
The connected four-point function is
\begin{align}
    G^{(4)}_c =&\frac{1}{4 N} \sigma_{w_1}^4 \bigg[ 3e^{4\sigma_{b_0}^2} e^{-\frac{\sigma_{w_0}^2}{2d}\sum_i x_i^2 -2(x_1 x_2 + \,6 \text{ perms})} \nonumber \\ &- (e^{-\frac{1}{2d} \sigma_{w_0}^2
    \left((x_{1}-x_{4})^2+(x_{2}-x_{3})^2\right)} + \,2 \text{ perms} 
    ) \bigg].
\end{align}
We see that the $2$-pt function is translation invariant, but not the $4$-pt function.

\medskip
\noindent \emph{Euclidean-Invariant Nets}. 
We are interested in constructing input layers that are sufficient to ensure invariance under $SO(d)$ and translations, i.e., under the Euclidean group. Consider an input layer of the form 
\begin{equation}
\ell_i(x) = F({\bf w^{(0)}}_i) \cos\left(\sum_j w^{(0)}_{ij} x_j + b^{(0)}_i\right), \,\,\, i \in {1,\dots,N},
\end{equation}
where the sum has been made explicit since $i$'s are not summed over, with 
\begin{equation}
    w^{(0)}_{ij} \sim P(w^{(0)}_{ij}) \qquad b^{(0)}_i \sim \text{Unif}[-\pi,\pi].
\end{equation}
It is easy to see that $x_i\to R_{ij} x_j$ for $R\in SO(d)$ can be absorbed into a redefinition of the $w^{(0)}_{ij}$'s, and the partition function is invariant under this action provided that $F({\bf w^{(0)}}_i)$ and $P(w^{(0)}_{ij})$ are invariant. Furthermore, arbitrary translations $x_j \to x_j+\epsilon_j$ give a term $w^{(0)}_{ij}\epsilon_j$ that can be absorbed into a redefinition of the $b^{(0)}_i$ that leaves the theory invariant. See \cite{Halverson:2021aot} for the $2$-pt and $4$-pt functions.

We emphasize two nice features:
\begin{itemize}
\item \textbf{Larger Euclidean Nets.} Any network that builds on $\ell$ without reusing its parameters is Euclidean-invariant.
\item \textbf{Spectrum Shaping.} In computing $G^{(2)}(p)$, $\textbf{b}_i$ gets evaluated on $p$, and $F$ may be chosen to shape the power spectrum (momentum space propagator) arbitrarily.
\end{itemize}
We refer the reader to \cite{Halverson:2021aot} for general calculations of $\ell$-correlators and to the specializations below for simple single-layer theories. We turn the $\ell_i(x)$ input layer into a scalar NN by
\begin{equation}
    \label{eqn:single_layer_Euclidean}
    \phi(x) = \sum w^{(1)}_i \ell_i(x)
\end{equation}
with drawn $w^{(1)}_i \sim P(w^{(1)})$.

\medskip
\noindent \emph{Cos-net.}
The architecture is defined by specifying \eqref{eqn:single_layer_Euclidean} to $F=1$ 
\begin{equation}
\phi(x) =  w^{(1)}_i \cos\left( w^{(0)}_{ij} x_j + b^{(0)}_i\right),
\end{equation}
and specific parameter densities
\begin{equation}
    w^{(1)} \sim \cN\left(0,\frac{\sigma_{w_1}^2}{N}\right) \qquad w^{(0)} \sim \cN\left(0,\frac{\sigma_{w_0}^2}{d}\right) \qquad b^{(0)} \sim \text{Unif}[-\pi,\pi].
\end{equation}
The two-point function is equivalent to that of Gauss-net
\begin{equation}
    G^{(2)}(x,y) = \frac{\sigma_{w_1}^2}{2} e^{-\frac{1}{2\din}\sigma_{w_0}^{2} (x_1-x_2)^2}.
\end{equation}
The connected four-point function is 
\begin{align}
    G^{(4)}|_{c} = &\frac{1}{8 N} \sigma_{w_1}^4 \bigg[ 3  \left(e^{-\frac{1}{2d} \sigma_{w_0}^2
   (x_{1}+x_{2}-x_{3}-x_{4})^2}+\, 2\text{ perms}\right) \nonumber \\ 
   &-\left(2 e^{-\frac{1}{2d} \sigma_b^2
   \left((x_{1}-x_{4})^2+(x_{2}-x_{3})^2\right)}+\, 2\text{ perms}\right) \bigg].
\end{align}
The $2$-pt function and $4$-pt function are fully Euclidean invariant.

Equipped with both Cos-net and Gauss-net, we'd like to compare the two:
\begin{itemize}
\item \textbf{Symmetry.} Cos-net is Euclidean-invariant $\forall N$, by construction, while Gauss-net enjoys this symmetry only when $N\to\infty$; in that limit it only depends on $G^{(2)}(x,y)$.
\item \textbf{Large-$N$ Duality.} As $N\to\infty$, both theories are drawn from the same Gaussian Process, i.e., with the same $G^{(2)}(x,y)$.
\end{itemize}

\medskip
\noindent \emph{Scalar-net.}
Here is one final example you might like, specializing \eqref{eqn:single_layer_Euclidean}
\begin{equation}
    \label{eqn:scalar-net}
    \phi(x) =  \sqrt{\frac{2 \text{vol}(B_\Lambda^d)}{(2\pi)^d \sigma_{w_1}^2}} \frac{1}{\sqrt{{\bf w^{(0)}}_i^2 + m^2}} \,\,w^{(1)}_i \cos\left( w^{(0)}_{ij} x_j + b^{(0)}_i\right),
\end{equation}
and specific parameter densities
\begin{equation}
    \label{eqn:scalar-net-densities}
    w^{(1)} \sim \cN\left(0,\frac{\sigma_{w_1}^2}{N}\right) \qquad w^{(0)} \sim \text{Unif}(B_\Lambda^d) \qquad b^{(0)} \sim \text{Unif}[-\pi,\pi],
\end{equation}
where $B_\Lambda^d$ is a $d$-ball of radius $\Lambda$. The theory is translation invariant by construction, and so we compute the power spectrum of the two-point function $G^{(2)}(x-y)$ to be 
\begin{equation}
G^{(2)}(p) = \frac{1}{p^2+m^2}.
\end{equation}
We see that we have a realization of the free scalar field theory in $d$ Euclidean dimensions. See Sec. \ref{sec:NNFT_phi4} for an extension to $\phi^4$ theory.

\section{Dynamics of Neural Networks}
\label{sec:dynamics}

Having covered expressivity and statistics, we turn to dynamics. Focusing on the most elementary NN dynamics, we ask
\Q{How does a NN evolve under gradient descent?} First we will study a simplification known as the neural tangent kernel (NTK), and then will use it in the case of MSE loss to solve a model exactly. We'll discuss drawbacks of the NTK, and then improve upon them with a scaling analysis that ensures feature learning.

We will study the dynamics of supervised learning with gradient  descent, with data 
\begin{equation}
\cD = \{(x_\alpha,y_\alpha)\}_{\alpha=1}^{|\cD|},
\end{equation}
and loss function
\begin{equation}
    \cL[\phi] = \frac{1}{|\cD|}\sum_{\alpha=1}^{|\cD|} \ell(\phi(x_\alpha), y_\alpha),
\end{equation}
We optimize the network parameters $\theta$ by gradient descent,
\begin{equation}
    \frac{d\theta_i}{dt} = -\eta \nabla_{\theta_i} \cL[\phi].
\end{equation}
It is also convenient to define 
\begin{equation}
\Delta(x) = - \frac{\delta\ell(\phi(x),y)}{\delta \phi(x)},
\end{equation}
where $y$ is to be understood as the label associated to $x$, which yields 
\begin{equation}
\label{eqn:gd_with_delta}
\frac{d\theta_i}{dt}= \frac{\eta}{|\cD|}\sum_{\alpha=1}^{|\cD|} \Delta(x_\alpha) \frac{\partial \phi(x_\alpha)}{\partial \theta_i}.
\end{equation}
as another form of the gradient descent equation, by chain rule. $\Delta(x)$ is the natural object of gradient descent in function space.

\textbf{We use Einstein summation throughout this section unless stated otherwise (which will happen).}

\subsection{Neural Tangent Kernel}

We now arrive at a classic dynamical result in ML theory, the neural tangent kernel \cite{NTK}. We train the network by gradient descent 
\begin{align}
\frac{d\phi(x)}{dt} &= \frac{\partial\phi(x)}{\partial\theta_i} \frac{d\theta_i}{dt} \\ &= \frac{\eta}{|\cD|}\sum_{\alpha=1}^{|\cD|} \Delta(x_\alpha) \,\Theta(x,x_\alpha),
\end{align}
where 
\begin{equation}
    \Theta(x,x_\alpha) = \frac{\partial\phi(x)}{\partial\theta_i}  \frac{\partial\phi(x_\alpha)}{\partial\theta_i}
\end{equation}
is the \emph{neural tangent kernel} (NTK).

Given the short derivation, this is clearly a fundamental object, but it seems terrible to work with, because it is:
\begin{itemize}
\item \textbf{Parameter-dependent.} For modern NNs, there are billions of parameters to sum over.
\item \textbf{Time-dependent.} Of course, the learning trajectory is $\theta(t)$, and therefore the NTK time evolves.
\item \textbf{Stochastic.} Since $\theta(t)$ begins at $\theta(0)$ sampled at initialization, the NTK inherits the randomness.
\end{itemize}
It's also non-local, communicating information about the loss at train points $x_\alpha$ to the test point $x$. In summary, the NTK is unwieldy.

The reason it is a classic result, however, is that it simplifies in the $N\to \infty$ limit. In this limit, neural network training is in the so-called
\begin{equation}
\text{Lazy regime:} \qquad |\theta(t) - \theta(0)| \ll 1,
\end{equation}
i.e. the number of parameters is large but the evolution keeps them in a local neighborhood. In such a regime the network is approximately a linear-in-parameters model \cite{NTK,lee2019wide}
\begin{equation}
\lim_{N\to \infty} \phi(x)\simeq \phi_\text{lin}(x) := \phi_{\theta_0}(x) + (\theta-\theta_0)_i \frac{\partial\phi(x)}{\partial\theta_i}\bigg|_{\theta_0},
\end{equation}
and we have 
\begin{equation}
    \lim_{N\to \infty}\Theta(x,x') \simeq \Theta(x,x')\big|_{\theta_0}.
\end{equation}
That is, the infinite-width NTK is the NTK at initialization, provided that the network evolves as a linear model. Furthermore, in the same limit the law of large numbers often allows a sum to be replaced by an expectation value, e.g., 
\begin{equation}
\lim_{N\to \infty} \Theta(x,x')|_{\theta_0} = \bk{\beta_\theta(x,x')} =: \bar \Theta(x,x'),
\end{equation}
for computable $\beta(x,x')$, yielding network dynamics governed by 
\begin{equation}
    \label{eqn:dynamics_frozen_NTK}
    \frac{d\phi(x)}{dt}= -\frac{\eta}{|\cD|}\sum_{\alpha=1}^{|\cD|} \frac{\delta l(\phi(x_\alpha),y_\alpha)}{\delta\phi(x_\alpha)} \,\, \bar \Theta(x,x_\alpha),
\end{equation}
where $\bar \Theta$ is the so-called \emph{frozen NTK}, a kernel that may be computed at initialization and fixed once-and-for-all. This is a dramatic simplification of the dynamics.

However, you should also complain.
\CT{Complaint}{The dynamics in \eqref{eqn:dynamics_frozen_NTK} simply interpolates between information at train points $x_\alpha$ and test point $x$,}
according to a fixed function $\bar \Theta$. This isn't ``learning" in  the usual NN sense, and there are \emph{zero} parameters. In particular, since the NN only affects \eqref{eqn:dynamics_frozen_NTK} through $\bar \Theta$, which is fixed, nothing happening dynamically in the NN is affecting the evolution. We say that in this limit the NN \textbf{does not learn features} in the hidden dimensions (intermediate layers), since their non-trivial evolution would cause the NTK to evolve.

\bigskip
\noindent \textbf{Example.} 
Let's compute the frozen NTK for a single-layer network, to get the idea. The architecture is 
\begin{equation}
    \phi(x) = \frac{1}{\sqrt{N}} \sum_{i=1}^N \sum_{j=1}^d w_i^{(1)} \sigma(w_{ij}^{(0)} x_j).
\end{equation}
We make the sums explicit here because one is very important.
The NTK is 
\begin{align}
\Theta(x,x') &= \sum_{i} \frac{\partial \phi(x)}{\partial w_i^{(1)}} \frac{\partial \phi(x')}{\partial w_i^{(1)}} + \sum_{ij} \frac{\partial \phi(x)}{\partial w_{ij}^{(0)}} \frac{\partial \phi(x')}{\partial w_{ij}^{(0)}} \\ 
&= \frac{1}{N} \sum_{i=1}^N \bigg(\sum_{j,l=1}^d \sigma(w_{ij}^{(0)}x_j)\sigma(w_{il}^{(0)}x'_l)\\ &\qquad + \sum_{j=1}^d x_j x'_j \,w_i^{(1)}w_i^{(1)}\, \sigma'(w_{ij}^{(0)} x_j) \sigma'(w_{ij}^{(0)}x'_j) \bigg) \\ &=: \frac{1}{N} \sum_{i }\beta_i(x,x').
\end{align}
If you squint a little, you'll see that the $i$-sum is a sum over the same type of object, $\beta_i(x,x')$, whose $i$ dependence comes from all these i.i.d. parameter draws in the $i$-direction. By the law of large numbers, we have that in the $N\to\infty$ limit
\begin{equation}
\bar \Theta(x,x') = \bk{\beta_i(x,x')},
\end{equation}
with no sum on $i$.
We emphasize 
\O{The NTK in the $N\to \infty$ limit is deterministic (parameter-independent), depending only on $P(\theta)$.}
Sometimes, the expectation may be computed exactly, and one knows the NTK that governs the dynamics once and for all.

\subsection{An Exactly Solvable Model}

Let us consider a special case of frozen-NTK dynamics with MSE loss,
\begin{equation}
    \ell(\phi(x),y) = \frac{1}{2}(\phi(x)-y)^2.
\end{equation}
Then the dynamics \eqref{eqn:dynamics_frozen_NTK} becomes 
\begin{equation}
    \frac{d\phi(x)}{dt} = -\frac{\eta}{|\cD|}\sum_{\alpha=1}^{|\cD|} (\phi(x_\alpha)-y_\alpha) \bar \Theta(x,x_\alpha).
\end{equation}
The solution to this ODE is 
\begin{equation}
    \phi_t(x) = \phi_0(x) + \frac{1}{|\cD|}\bar \Theta(x,x_\alpha) \bar\Theta(x_\alpha, x_\beta)^{-1}\left(\mathbbm{1}-e^{-\eta \bar\Theta t}\right)_{\beta \gamma}\left(y_\gamma-\phi_0(x_\gamma)\right),
\end{equation}
where computational difficulty is that $\bar\Theta(x,x_\alpha)$ is a $|\cD|\times |\cD|$ matrix and takes $O(|\cD|^3)$ time to invert. 
The solution defines a trajectory through function space from $\phi_0$ to $\phi_\infty$. The converged network is 
\begin{equation}
    \phi_\infty(x) = \phi_0(x) +\bar \Theta(x,x_\alpha) \bar\Theta(x_\alpha, x_\beta)^{-1}\left(y_\beta-\phi_0(x_\beta)\right).
\end{equation}
This is known as $\textbf{kernel regression}$, a classic technique in ML. 
In general kernel regression, one chooses the kernel. In our case, gradient descent training in the $N\to \infty$ limit \emph{is} kernel regression, with respect to a specific kernel determined by the NN, the NTK $\bar \Theta$. 

On train points we have \textbf{memorization}
\begin{equation}
    \phi_\infty(x_\alpha) = y_\alpha \qquad \forall \alpha.
\end{equation}
On test points $x$, the converged network is performing an interpolation, communicating residuals $R_\beta$ on train points $\beta$ through a fixed kernel $\bar\Theta$ to test points $x$. The prediction depends on $\phi_0$, but may be averaged over to obtain 
\begin{equation}
\mu_\infty(x) := \bk{\phi_\infty(x)} = \bar\Theta(x,x_\alpha) \bar\Theta(x_\alpha, x_\beta)^{-1}y_\beta,
\end{equation}
provided that $\bk{\phi_0}=0$, as in many initializations for the parameters.
Let's put some English on the \textbf{remarkable facts},
\begin{itemize}
    \item $\mu_\infty(x)$ is the mean prediction of an $\infty$ number of $\infty$-wide NNs trained to $\infty$ time.
    \item  If $\phi_0$ is drawn from a GP, then $\phi_\infty$ is as well. The mean is precisely $\mu_\infty(x)$, see \cite{lee2019wide} for the two-point function and covariance.
\end{itemize}

\subsection{Feature Learning}

The frozen NTK is a tractable toy model, but it has a major drawback: it does not learn features.
In this section we perform a more general study of learning dynamics, with a focus on choosing wise $N$-scaling such that features are non-trivially learned during gradient descent. To do this we will engineer three properties: that features (pre-activations) are finite at initialization, predictions evolve in finite time, and features evolve in finite time. 

This section is denser than the others, so I encourage the reader to remember that we are aiming to achieve these three principles if they become overwhelmed with details. Throughout, I am following lecture notes of Pehlevan and Bordelon \cite{pehlevan2024lecture}  with some changes in notation to match the rest of the lectures; see those lectures for more details, \cite{yang2020feature,bordelon2022self} for original literature from those lectures that I utilize, and \cite{roberts2022principles,yaida2022meta} for related work on feature learning.

We study a deep feedforward network with $L$ layers and width $N$, which in all is a map
\begin{equation}
\phi:\bR^D \to \bR
\end{equation}
(note the input dimension $D$; $d$ is reserved for below) defined recursively as 
\begin{align}
\phi(x) &= \frac{1}{\gamma_0 N^d}z^{(L)}(x) \\ 
z^{(L)}(x) &= \frac{1}{N^{a_L}} w_i^{(L)}\, \sigma(z_i^{(L-1)}(x))\\
z^{(\ell)}_i(x) &= \frac{1}{N^{a_\ell}} W^{(\ell)}_{ij} \sigma(z_j^{(\ell-1)}(x)) \\
z_i^{(1)}(x) &= \frac{1}{N^{a_1}\sqrt{D}} W^{(1)}_{ij}x_j
\end{align}
where Einstein summation is implied throughout this section   (unless stated otherwise) and all Latin indices run from $\{1,\dots,N\}$ \emph{except in the $j$-index in the first layer}, when they are $\{1,\dots,D\}$. The parameters are drawn 
\begin{equation}
w_i^{(\ell)} \sim \cN\left(0,\frac{1}{N^{b_L}}\right) \qquad W^{(\ell)}_{ij} \sim \cN\left(0,\frac{1}{N^{b_\ell}}\right).
\end{equation}
We scale the learning rate as 
\begin{equation}
\eta = \eta_0 \gamma_0^2 N^{2d-c}
\end{equation}
with $\gamma_0,\eta_0$ $O(1)$  constants. We use a parameterization that will be convenient where $d$ has already been introduced but $c$ is a new parameter. For notational brevity we will sometimes use a Greek index subscript in place of inputs, e.g. $z^{(\ell)}_\alpha:= z^{(\ell)}(x_\alpha)$. The $z$'s are known as the pre-activations, as they are the inputs to the activation functions $\sigma$.

We have a standard MLP but have parameterized our ignorance of $N$-scaling, governed by parameters $(a_\ell,b_\ell,c,d)$. We will use this freedom to set some reasonable \textbf{goals}:
\begin{itemize}
    \item \textbf{Finite Initialization Pre-activations.} $z^{(\ell)} \sim O_N(1)\,\,\,\forall l$.
    \item \textbf{Learning in Finite Time.} $d\phi(x)/dt\sim O_N(1)$.
    \item \textbf{Feature Learning in Finite Time.} $dz^{(\ell)}/dt \sim O_N(1) \,\,\, \forall l$.
\end{itemize}
These constraints have a one-parameter family of solutions, which becomes completely fixed under an additional learning rate assumption. We take each constraint in turn.

\bigskip
\noindent \emph{Finite Pre-activations.} Since the weights are zero mean,  trivially
\begin{equation}
\bk{z^{(\ell)}_\alpha} = 0 \qquad \forall l.
\end{equation}
We must also compute the covariance. For the first pre-activation, we have 
\begin{align}
\bk{z_{i\alpha}^{(1)}z_{j\beta}^{(1)}} = \frac{1}{D N^{2a_1}} \bk{w^{(1)}_{im}w^{(1)}_{jn}} \, x_{m\alpha}x_{n\beta} = \delta_{ij}\frac{1}{DN^{2a_1+b_1}} x_{m\alpha}x_{m\beta}.
\end{align}
A similar calculation for higher layers $l$  gives
\begin{align}
    \bk{z_{i\alpha}^{(\ell)}z_{j\beta}^{(\ell)}} &= \frac{1}{ N^{2a_\ell}} \bk{w^{(1)}_{im}w^{(1)}_{jn}} \, \bk{\sigma(z^{(\ell-1)}_{m\alpha})\sigma(z^{(\ell-1)}_{m\beta})} \\ &= \delta_{ij}\frac{1}{N^{2a_\ell+b_\ell-1}} \frac{1}{N} \bk{\sigma(z^{(\ell-1)}_{m\alpha})\sigma(z^{(\ell-1)}_{m\beta})} \\
    &= \delta_{ij}\frac{1}{N^{2a_\ell+b_\ell-1}} \bk{\Phi^{(\ell-1)}_{\alpha\beta}},
\end{align}
where 
\begin{equation}
    \Phi^{(\ell)}_{\alpha\beta} := \frac{1}{N} \sigma(z^{(\ell)}_{m\alpha})\sigma(z^{(\ell)}_{m\beta})
\end{equation}
is a \emph{feature kernel}. We are constructing a proof-by-induction that the pre-activations are $O_N(1)$, so at this stage we may assume that the pre-activations $z^{(\ell-1)} \sim O_N(1)$ and therefore $\Phi^{(\ell-1)} \sim O_N(1)$ since it is the average of $N$ $O_N(1)$ quantities $\sigma(z^{(\ell-1)}_{m\alpha})$. With this in hand, 
\begin{equation}
2a_1+b_1=0 \qquad 2a_\ell + b_\ell = 1 \,\, \forall l > 1,
\end{equation}
ensures that the pre-activations $z^{(\ell)}$ are $O_N(1)$, as is empirically required for well-behaved training.

As an aside: in the $N\to \infty$ limit feature kernels asymptote to deterministic objects, akin to the frozen NTK behavior we have already seen, and intermediate layer pre-activations $z_\alpha^{(\ell)}$ are also Gaussian distributed. Therefore in that limit, the statistics of a randomly initialized neural network is described by a sequence of generalized free field theories where correlations are propagated down the network according to a recursion relation; see, e.g., \cite{schoenholz2016deep,lee2019wide}.

\bigskip
\noindent \emph{Learning in Finite Time.} Similar to our NTK derivation, we have 
\begin{equation}
\frac{d\phi(x)}{dt} = \frac{\eta}{|\cD|}\sum_{\alpha=1}^{|\cD|} \Delta(x_\alpha) \frac{\partial\phi(x_\alpha)}{\partial\theta_i}\frac{\partial\phi(x)}{\partial\theta_i},
\end{equation}
We have that 
\begin{equation}
\frac{d\phi(x)}{dt}\sim O_N(1)\,\,\, \leftrightarrow \,\,\,\frac{\gamma_0^2 N^{2d}}{N^c} \frac{\partial\phi(x_\alpha)}{\partial\theta_i}\frac{\partial\phi(x)}{\partial\theta_i} \sim O_N(1).
\end{equation}
A short calculation involving a chain rule, keeping track of the different types of layers, and tedium yields
\begin{align}
    \frac{\eta_0\gamma_0^2 N^{2d}}{N^c} \frac{\partial\phi(x_\alpha)}{\partial\theta_i}\frac{\partial\phi(x)}{\partial\theta_i}
&= \frac{1}{N^c} \bigg[ \frac{1}{N^{2 a_L - 1}} \Phi_{\mu \nu}^{(L-1)} \nonumber \\ &+ \sum_{\ell=2}^{L-1} \frac{1}{N^{2 a_\ell - 1}} G_{\alpha x}^{(\ell)} \Phi_{\alpha x}^{(\ell-1)} + \frac{1}{N^{2 a_1}} G_{\alpha x}^{(1)} \Phi_{\alpha x}^{(0)} \bigg]
\end{align}
where 
\begin{equation}
G_{\mu\nu}^{(\ell)} = \frac{1}{N} g^{(\ell)}_{i\mu}\, g^{(\ell)}_{i\nu},
 \qquad g^{(\ell)}_{i\mu} = \sqrt{N}\, \frac{\partial z^{(L)}_\mu}{\partial z^{(\ell)}_{i\mu}} 
\end{equation}
arises naturally when doing a chain rule through the layers and $x$ in the subscript is a stand-in for an arbitrary input $x$. Some additional work \cite{pehlevan2024lecture} shows that $g, G \sim O_N(1)$. Since we've shown that the feature kernels are also $O_N(1)$, then if 
\begin{equation}
    \label{eqn:constraints_learning_predictions}
2a_1 + c = 0, \qquad 2a_\ell + c = 1 \,\,\, \forall l > 1,
\end{equation}
we have that $\frac{d\phi(x)}{dt}\sim O_N(1)$, i.e., predictions evolve in finite time, and therefore learning happens in finite time.

\bigskip
\noindent \emph{Features Evolve in Finite Time.} 
Finally, we need the features to evolve. For the first layer we compute
\begin{align} 
 \frac{dz_{i\mu}^{(1)}}{dt}&=\frac{1}{N^{a_1}\sqrt{D}} \frac{dW_{ij}}{dt} x_{j\mu} 
 = \frac{1}{N^{2a_1+c-d+1/2}}  \frac{\eta_0\gamma_0}{|\cD|} 
 \sum_{\nu=1}^{|\cD|}\Delta_\nu g_{i\nu}^{(1)} \Phi_{\mu\nu}^{(0)},
\end{align} 
where the only $N$-dependence arises from the exponent. Given \eqref{eqn:constraints_learning_predictions}, 
\begin{equation}
d=\frac12
\end{equation}
is required to have  $dz^{(1)}/dt \sim O_N(1)$, a constraint that in fact persists for all $dz^{(\ell)}/dt$. We note that we have
\CT{No feature learning}{if $d< \frac12$,}
which applies in particular to the NTK regime $d=0$.

\bigskip
\noindent \emph{Summarizing the Constraints}. Putting it all together, by enforcing finite pre-activations and that predictions and features evolve in finite time, we have constraints given by
\begin{align}
    2a_1 + b_1 &= 0 \\
    2a_\ell + b_\ell &= 1 \quad \forall l > 1 \\
    2a_1 + c &= 0 \\
    2a_\ell + c &= 1 \quad \forall l > 1 \\
    d &= \frac12.
\end{align}
These constraints are solves by a one-parameter family depending on $a\in \bR$
\begin{align}
(a_\ell, b_\ell, c_\ell) &= (a,1-2a,1-2a) \quad \forall l > 1 \\ 
(a_1, b_1, c_1) & = (a-\frac12, 1-2a, 1-2a).
\end{align}
If one makes the addition demands that $\eta \sim O_N(N^{2d-c})$ is $O_N(1)$, so that you don't have to change the learning rate as you scale up, then $c=1$ is fixed and there is a unique solution. This scaling is known as the \emph{maximal update parameterization}, or $\mu P$ \cite{yang2020feature}.

\section{NN-FT Correspondence}
\label{sec:NNFT}

Understanding the statistics and dynamics of NNs has led us naturally to objects that we are used to from field theory. The idea has been to understand ML theory, but one can also ask the converse, whether ML theory gives new insights into field theory. With that in mind, we ask
\Q{What is a field theory?}
At the very least, a field theory needs \begin{itemize}
    \item \textbf{Fields}, functions from an appropriate function space, or sections of an appropriate bundle, more generally. 
    \item \textbf{Correlation Functions} of fields, here expressed as scalars
    \begin{equation}
        G^{(n)}(x_1,\dots,x_n) = \bk{\phi(x_1)\dots \phi(x_n)}.
    \end{equation}
\end{itemize}
You might already be wanting to add more beyond these minimal requirements -- we'll discuss that in a second. For now, we have 
\CT{Answer}{a FT is an ensemble of functions with a way to compute their correlators.}
In the Euclidean case, when the expectation is a statistical expectation, one my say
\CT{Euclidean Answer}{a FT is a statistical ensemble of functions.}
Our minimal requirements get us a partition function
\begin{equation}
Z[J] = \bk{e^{\int d^dx J(x)\phi(x)}}
\end{equation}
that we can use to compute correlators, where at this stage we are agnostic about the definition of $\bk{\cdot}$. In normal field theory, the $\bk{\cdot}$ is defined by the Feynman path integral 
\begin{equation}
Z[J] = \int \cD \phi \, e^{-S[\phi] + \int d^dx J(x)\phi(x)},
\end{equation}
which requires specifying an action $S[\phi]$ that determines a density on functions $\exp(-S[\phi])$. But that's not the data we specify when we specify a NN. The NN data $(\phi_\theta, P(\theta))$ instead defines 
\begin{equation}
Z[J] = \int d\theta P(\theta) e^{\int d^dx J(x)\phi_\theta(x)}.
\end{equation}
These are two different ways of defining a field theory, and indeed given $(\phi_\theta,P(\theta))$ one can try to work out the associated action, in which case we have dual description of the same field theory, as in the NNGP correspondence.
The parameter space description is already quite useful, though, as it enables the computation of correlation functions even if the action isn't known. In certain cases it enables the computation of exact correlators in interacting  theories.

\medskip
Okay, you get it, this is a different way to do field theory. Now I'll let you complain about my definition. You're asking 
\Q{Shouldn't my definition of field theory include $X$?}
I'm writing this before I give the lecture, and my guess is you already asked about a set of $X$'s, e.g. 
\begin{align}
X \in \{
    \text{Quantum} ,
    \text{ Lagrangian},
    \text{ Symmetries},
    \text{ Locality}, \dots
    \}.
\end{align}
The problem is that with any such $X$, there's usually some community of physicists that doesn't care. For instance, not all statistical field theories are Wick rotations of quantum theories; not all field theories have a known Lagrangian; not all field theories have symmetry; not all field theories are local. So I'm going to stick with my definition, because at a minimum I want fields and correlators.

Instead, if your $X$ isn't included in the definition of field theory, it becomes an engineering problem. Whether you're defining your specific theory by $S[\phi]$, $(\phi_\theta,P(\theta))$, or something else, you can ask
\Q{Can I engineer my defining data to get FT + $X$?}
For $X=\text{Symmetries}$ you've already seen ways to do this at the level of actions in QFT1 and at the level of $(\phi_\theta,P(\theta))$ in these lectures.

For a current account of recent progress in NN-FT, see the rather long Introduction of \cite{Demirtas:2023fir} and associated references, as well as results.

\subsection{Quantum Field Theory}

We've been on Euclidean space the whole time\footnote{This can be relaxed, see e.g. for a recent paper defining an equivariant network in Lorentzian signature \cite{zhdanov2024clifford}.}, so it's natural to wonder in what sense these field theories are \emph{quantum}. In a course on field theory, we first learn to canonically quantize and then at some later point learn about Wick rotation, and how it can define Euclidean correlators. The theory is manifestly quantum.

But given a Euclidean theory, can it be continued to a well-behaved quantum theory in Lorentzian signature, e.g. with unitary time evolution and a Hilbert space without negative norm states? If we have a nice-enough local action, it's possible, but what if we don't have an action? We ask:
\Q{Given Euclidean correlators, can the theory be continued to a well-behaved Lorentzian quantum theory?}
This is a central question in axiomatic quantum field theory, and the answer is that it depends on the properties of the correlators. The Osterwalder-Schrader (OS) theorem \cite{Osterwalder:1973dx} gives a set of conditions on the Euclidean correlators that ensure that the theory can be continued to a unitary Lorentzian theory that satisfies the Wightman axioms.
The conditions of the theorem include 
\begin{itemize}
\item \textbf{Euclidean Invariance}. The correlators are invariant under the Euclidean group, which after continuation to Lorentzian signature becomes the Poincar\'e group.
\item \textbf{Permutation Invariance} of the correlators $G^{(n)}(x_1,\dots,x_n)$ under any permutation of the $x_1,\dots,x_n$.
\item \textbf{Reflection Positivity}. Having time in Lorentzian signature requires picking a Euclidean time direction $\tau$. Let $R(x)$ be the reflection of $x$ in the $\tau=0$ plane. Then reflection positivity requires that
\begin{equation}
    G^{(2n)}(x_1,\dots,x_n,R(x_1),\dots,R(x_n)) \geq 0.
\end{equation}
Technically, this is necessary but not sufficient. An accessible elaboration can be found in notes \cite{Simmons-Duffin:2016gjk} from a previous TASI.
\item \textbf{Cluster Decomposition} occurs when the connected correlators vanish when any points are infinitely far apart.
\end{itemize}
If all of these are satisfied, then the pair $(\phi_\theta, P(\theta))$ that defines the NN-FT actually defines a neural network \emph{quantum} field theory \cite{Halverson:2021aot}.
In NN-FT, permutation invariance is essentially automatic and Euclidean invariance may be engineered as described in Section \ref{sec:stats_examples}. Cluster decomposition and reflection positivity hold in some examples \cite{Halverson:2021aot,Demirtas:2023fir}, but systematizing their construction is an important direction for future work.

There is at least one straightforward way to obtain an interacting NN-QFT. Notably, if $(\phi_\theta, P(\theta))$ is a NNGP that satisfies the OS axioms (this is much easier \cite{Halverson:2021aot}) with Gaussian partition function
\begin{equation}
Z_{G}[J] = \int d\theta P(\theta) e^{\int d^dx J(x)\phi_\theta(x)},
\end{equation}
then one may insert an operator associated to any local potential $V(\phi)$, which deforms the action in the expected way and the NN-FT to 
\begin{align}
Z[J] &= \int d\theta P(\theta) e^{ \int d^dx V(\phi_\theta(x))} e^{\int d^dx J(x)\phi_\theta(x)} \\
&=: \int d\theta \tilde P(\theta) e^{\int d^dx J(x)\phi_{\theta}(x)},
\end{align}
where the architecture equation $\phi_\theta(x)$ lets us sub out the abstract expression for a concrete function of parameters, defining a new density on parameters $\tilde P(\theta)$ in the process. The interactions in $V(\phi)$ break Gaussianity of the NNGP that was ensured by a CLT. This means a CLT assumption must be violated: it is the breaking of statistical independence in $\tilde P(\theta)$. The theory $Z[J]$ defined by $(\phi_\theta, \tilde P(\theta))$ is an interacting NN-QFT, since local potentials that deformed Gaussian QFTs still satisfy reflection positivity and cluster decomposition.

\subsection{$\phi^4$ Theory}
\label{sec:NNFT_phi4}
With this discussion of operator insertions, it's clear now how to get $\phi^4$ theory. We just insert the operator
\begin{equation}
e^{\int d^dx \phi_\theta(x)^4}
\end{equation}
into the partition function associated to the free scalar of Section \ref{sec:stats_examples}; this operator with the architecture \eqref{eqn:scalar-net} technically requires an IR cutoff, though other architectures realizing the free scalar may not. The operator insertion deforms the parameter densities in \eqref{eqn:scalar-net-densities} and breaks their statistical independence, explaining the origin of interactions in the NN-QFT. See \cite{Demirtas:2023fir} for a thorough presentation.

\subsection{Open Questions}

In this section we've discussed a neural network approach to field theory in which the partition function is an integral over probability densities of parameters, and the fields (networks) are functions of these parameters. We have summarized some essential results, but there are many outstanding questions:
\begin{itemize}
\item \textbf{Reflection Positivity}. Is there a systematic way to engineer NN-FTs $(\phi_\theta, P(\theta))$ that satisfy reflection positivity?
\item \textbf{Cluster Decomposition}. Can  systematically understand the conditions under which cluster decomposition holds in NN-FT?
\item \textbf{Engineering Actions}. How generally can we define an NN-FT that is equivalent to a field theory with a fixed action?
\item \textbf{Locality} in the action can be realized at infinite-$N$, but is there a fundamental obstruction at finite-$N$?
\end{itemize}
On the other hand, an NN-FT approach to CFT and to Grassmann fields is well underway and should appear in 2024.

\section{Recap and Outlook}

Neural networks are the backbone of recent progress in ML. In these lectures we took the perspective that to understand ML, we must understand neural networks, and built our study around three pillars: the expressivity, statistics, and dynamics of neural networks.

Let's recap the essentials and then provide some outlook. In Section \ref{sec:expressivity} on expressivity, we asked ``How powerful is a NN?" We presented the Universal Approximation Theorem (UAT) and gave a picture demonstrating how it works. We also presented the Kolmogorov-Arnold Representation Theorem (KART), which recently motivated a new architecture for deep learning where activations are learned and live on the edges. We wonder:
\CT{Outlook}{Cybenko's UAT was around for over 20 years before the empirical breakthroughs of the 2010's. Existence of good approximators wasn't enough, we needed better compute and optimization schemes to find them dynamically.  However, given tremendous progress in dynamics, should we systematically return to the approximation theory literature, to help motivate new architectures, as e.g., in KART and KAN?}

In Section \ref{sec:statistics} on statistics, I reminded the reader that a randomly initialized NN is no more fundamental than a single role of the dice. It is a random function with parameters that is drawn from a statistical ensemble of NNs, and we asked ``What characterizes the statistics of the NN ensemble?'' I reviewed a classic result of Neal from the 90's that demonstrated that under common assumptions a width-$N$ feedforward network is drawn from a Gaussian Process --- a Gaussian density on functions --- in the $N\to \infty$ limit. I explained that work from the last decade has demonstrated that this result generalizes to many different architectures, as a result of the Central Limit Theorem. Non-Gaussianities therefore arise by violating an assumption of the Central Limit Theorem, such as $N\to \infty$ or statistical independence, and by violating them weakly the non-Gaussianities can be made parametrically small. In field theory language, NNGPs are generalized free field theories and NN Non-Gaussian processes are interacting field theories. I also introduced a mechanism for global symmetries and exemplified many of the phenomena in the section.

In Section \ref{sec:dynamics} I presented some essential results from ML theory on NN dynamics, asking ``How does a NN evolve under gradient descent?'' When trained with full-batch gradient descent and a particular normalization, the network dynamics are governed by the Neural Tangent Kernel (NTK) that in general is an intractable object. However, in $N\to \infty$ limits it becomes a deterministic function and the NTK at initialization governs the dynamics for all time. An exactly solvable model with mean-squared error loss is presented that includes the mean prediction of an infinite number of infinitely wide neural networks trained to infinite time. The mean network prediction is equivalent to kernel regression, and the kernel communicates information from train points to test points in order to make predictions. However, from this description of the NTK a problem is already clear: nothing is being learned, and in particular late-time features in the hidden dimensions are in a local neighborhood of their initial values.  I showed how a detailed $N$-scaling analysis allows one to demand that network features and predictions update non-trivially, leading to richer learning regimes known as dynamical mean field theory or the maximal update parameterization.
\CT{Outlook}{These recent theories of NN statistics and dynamics sometimes realize interesting toy models, but with known shortcomings from a learning perspective, though the $N$-scaling analysis in the feature learning section is promising. Surely there is still much to learn overall, e.g., we have said very little in these lectures about architecture design principles. For instance, some architectures, such as the Transformers \cite{vaswani2017attention} central to LLMs, are motivated by structural principles, whereas others are motivated by theoretical guarantees. \vspace{.5cm} \\ What are the missing principles that combines statistics, dynamics, and architecture design to achieve optimal learning?}

Finally, in Section \ref{sec:NNFT} I explained how neural networks provide a new way to define a field theory, so that one might use ML theory for physics and not just physics for ML theory. This NN-FT correspondence is a rethink of field theory from an ML perspective, and I tried to state clearly about what is and isn't known about obtaining cherished field theory properties in that context. For example, I explained how by engineering a desired NNGP, such as the free scalar, one may do an operator insertion in the path integral and interpret the associated interactions as breaking the statistical independence necessary for the CLT. By such a mechanism, one may engineer $\phi^4$ theory in an ensemble of infinite width neural networks.
\CT{Outlook}{Can NN-FTs motivate new interesting physics theories or provide useful tools for studying known theories?}

Our discussion in this Section is directly correlated with the topics of these lectures. It reviews recent theoretical progress, but the careful reader surely has a sense that theory is significantly behind experiment and there is still much to do. It is a great time to enter  the field, and to that end I recommend carefully reading the literature that I've cited, as well as the book \emph{Geometric Deep Learning} \cite{bronstein2021geometric} that covers many current topics that will be natural to a physicist, including grids, groups, graphs, geodesics, and gauges. Many aspects of that book are related to principled architecture design, which complements the statistical and dynamical approach that I've taken in these lectures. I also recommend the book \emph{Deep Learning} \cite{GoodBengCour16} for a more comprehensive introduction to the field.

I end with a final analogy between history and the current situation in ML. We are living in an ML era akin to the pre-PC era in classical computing. Current large language models (LLMs) or other deep networks are by now very powerful, but are the analogs of room-sized computers that can only be fully utilized by the privileged. Whereas harnessing the power of those computers in the 60's required access to government facilities or large laboratories, only billion dollar companies have sufficient funds and compute to train today's LLMs from scratch. Just as the PC revolution brought computing power to the masses in the 80's, a central question now is whether we can do the analog in ML by learning how to train equally powerful small models with limited resources. Doing so likely requires a deeper understanding of theory, especially with respect to sparse networks.

\vspace{.5cm}
\noindent \textbf{Acknowledgements:} I would like to thank the students and organizers of TASI for an inspiring scientific environment and many excellent questions. I am grateful for the papers of and discussions with friends whose works have contributed significantly to these lectures, especially Yasaman Bahri, Cengiz Pehlevan, and Greg Yang.  I'd also like to thank my collaborators on related matters, including Mehmet Demirtas, Anindita Maiti, Fabian Ruehle, Matt Schwartz, and Keegan Stoner. I would like to thank Sam Frank, Yikun Jiang, and Sneh Pandya for correcting typos in a draft. Finally, thanks to GitHub Copilot for its help in writing these notes, including the correct auto-completion of equations and creation of \texttt{tikz} figures! Two-column landscape mode was chosen to facilitate boardwork, but I think it also makes a pleasant read, HT @ \cite{ginsparg1988applied}. I am supported by the National Science Foundation under CAREER grant PHY-1848089 and
Cooperative Agreement PHY-2019786 (The NSF AI Institute for Artificial Intelligence and Fundamental
Interactions).

\vspace{.5cm}
\noindent \textbf{Disclaimers:} In an effort to post these notes shortly after TASI 2024, there are more typos and fewer references than is ideal. Other references and topics may be added in the future to round out the content that was presented in real-time. I also struck a playful tone throughout, because lectures are supposed to be fun, but may have overdone it at times. I welcome suggestions on the above, but updates will still aim for clarity and brevity, targeted at HET students.

\appendix 
\section{Central Limit Theorem}
\label{app:CLT}
Let us recall a simple derivation of the Central Limit Theorem (CLT), in order to better understand the statistics of neural networks. Consider a sum of random variables
\begin{equation}
\phi = \frac{1}{\sqrt{N}} \sum_{i=1}^N X_i,
\end{equation}
with $\bk{X_i} = 0$. The moments $\mu_r$ and cumulants $\kappa_r$ are determined by the moment generating function (partition function) $Z[J] = \bk{e^{J\phi}}$ and cumulant generating function $W[J] = \log Z[J]$, respectively, as 
\begin{align}
    \mu_r & = \left(\frac{d}{dJ}\right)^r Z[J]\bigg|_{J=0} \\
    \kappa_r & = \left(\frac{d}{dJ}\right)^r W[J]\bigg|_{J=0}. 
\end{align}
If the $X_i$ are independent random variables, then the partition function factorizes 
$Z_{\sum_i X_i}[J] = \prod_i Z_{X_i}[J]$, and the cumulant generating function of the sum is the sum of the cumulant generating functions, yielding
\begin{align}
    W_{\sum_i X_i}[J] = \sum_i W_{X_i}[J] \label{eqn:Windep}\\
    \kappa_r^{\sum X_i} = \sum_i \kappa_r^{X_i}.
\end{align} 
If the $X_i$ are identically distributed, then the cumulants $\kappa_r^{X_i}$ are the same for all $i$ and we account for the $1/\sqrt{N}$ appropriately, we obtain
\begin{equation}
\kappa_r^\phi = \frac{\kappa_r^{X_i}}{N^{r/2-1}}.
\end{equation}
This yields 
\begin{equation}
\lim_{N\to\infty} \kappa_{r>2}^\phi = 0, \label{eqn:CLT_kappar}
\end{equation}
which is sufficient to show that $\phi$ is Gaussian in the large-$N$ limit. In physics language, cumulants are connected correlators, and \eqref{eqn:CLT_kappar} means that Gaussian (free) theories have no connected correlators.

In neural networks we will be interested in studying certain Gaussian limits. From this CLT derivation, we see two potential origins of non-Gaussianity:
\begin{itemize}
    \item \textbf{$1/N$-corrections} from appearance in $\kappa_r^\phi$.
    \item \textbf{Independence breaking} since the proof relied on \eqref{eqn:Windep}.
\end{itemize}

\bibliographystyle{utphys}
\bibliography{refs}

\end{document}